\documentstyle[12pt,epsf]{article}
\begin{document}

\title{{\Large {\bf The Phase Structure of Antiferromagnetic Ising Model in the
Presence of Frustrations}}}
\author{A.Z. Akheyan\thanks{
e-mail: akheyan@lx2.yerphi.am} ,N.S. Ananikian\thanks{
e-mail: ananik@jerewan1.yerphi.am} , S.K. Dallakian\thanks{
e-mail: saco@atlas.yerphi.am} \\
{\normalsize Department of Theoretical Physics, Yerevan Physics Institute,}\\
{\normalsize Alikhanian Br.2, 375036 Yerevan, Armenia}}
\maketitle

\begin{abstract}
The antiferromagnetic Ising model in a magnetic field is considered on the
Husimi tree. Using iteration technique we draw the plots of magnetization
versus external field for different temperatures and construct the resulting
phase diagram. We show that frustration effects essentially change the
critical properties. The model exhibits three distinct critical regions,
including re-entrant phase structure and Griffiths singularities.
\end{abstract}

\newpage

It is now widely recognized that Bethe-like approximations are more reliable
than conventional mean-field approximation. In mean-field approximation one
replaces the interaction of a given site with the rest of a system by a mean
field and thus completely neglects fluctuations. In Bethe approximation one
first calculates the fluctuations in a microscopic region and only then
replaces the interaction to the rest of a system by a mean field. For
ferromagnetic and unfrustrated antiferromagnetic models these both
approaches provide the same qualitative picture, since both have the same
symmetries and large scale behavior. However in models with frustrations the
microscopic fluctuations along a loop with even number of bond play
important role and essentially change critical picture. Such systems attract
now great attention from both experimental and theoretical point of view.
The most common example of these models are triangular antiferromagnets,
which were reviewed in detail in \cite{A1,A2}.

It is well known that instead of Bethe-like approximation one can consider
exact solution of the models on Bethe or Bethe like hierarchical lattices.
To obtain frustration effect we use so called Husimi tree, which is in fact
the Bethe lattice constructed from triangles (fig. 1). We observe the most
simple Ising model on Husimi tree defined by Hamiltonian:

\begin{equation}
\label{R1}-H/kT=J\sum_{(i,j)}{\sigma }_i{\sigma }_j+h\sum_i{\sigma }_i, 
\end{equation}
with antiferromagnetic coupling ($J<0$). We have constructed the plots of
magnetization versus external field for different temperature and drawn the
resulting phase diagram. Due to frustration effects this diagram essentially
differs from the one obtained on Bethe lattice. Moreover we observed new
critical region which, to our knowledge, has never been described in
literature.

Husimi tree can be built successively by attaching $\gamma $ new triangles
to each free site of a previous shell. According to this construction one
can present the site magnetization $m$ $=<\sigma >$ by equation\cite{A3}:

\begin{equation}
\label{R2}m=\frac{e^{2h}x^\gamma -1}{e^{2h}x^\gamma +1}, 
\end{equation}
where $x$ is obtained from the attractor of the following map:

\begin{equation}
\label{R3}x_n=f(x_{n-1}),\qquad f(x)=\frac{e^{4(h+J)}x^{2(\gamma
-1)}+2e^{2h}x^{\gamma -1}+1}{e^{4h}x^{2(\gamma -1)}+2e^{2h}x^{\gamma
-1}+e^{4J}}, 
\end{equation}

The exact recursion formula allows us to study in detail the critical
properties. In thermodynamic limit ($n\rightarrow \infty $) the recursion
sequence $\left\{ x_n\right\} $ either converges to the stable fixed point $%
x^{*}$ or to the stable 2-cycle $(x_1$,$x_2)$ . First case corresponds to
the phase without sublattice structure, while in second case we have the
antiferromagnetic ordering where the values $m_1$ and $m_2$ are the
magnetizations of two sublattices\cite{S1}. Proceeding from these
principles, we can now construct the plots of magnetization versus external
field for different temperatures.

In fig.2 we present the plot of $m$ versus $h$ at low temperature. This
graphic is typical for antiferromagnetic models at low temperature and is
the result of competition between first and second term in Hamiltonian (\ref
{R1}). The plots of magnetization for higher temperatures are shown in
fig.3,4, and resulting phase diagram is shown in fig.5.

The graphics like shown in fig.3 are completely absent in antiferromagnetic
Ising model on Bethe lattice, and its appearance here is due to frustrations
on triangles.

The above result can be explained by considering the frustration as a
background disorder in the sense that if two spins on triangle aligned
antiparallel then the average value of magnetization on the third site is
zero in the absence of magnetic field at all temperatures. Note that the
background disorder due to frustration disappears when we introduce the
magnetic field.

Thus when we increase the temperature, the background disorder makes
spontaneous magnetization zero at lower temperature $T_1$ than it should be
in unfrustrated case, and, as a result, one obtains the dependence of
magnetization from magnetic field shown in figure 3. This eight-like
structure survives until reaching to another temperature $T_2$. Note that at
all range of temperatures between $T_1$and $T_2$ we have continuous phase
transitions in zero magnetic field.

As we continue to increase temperature the non zero magnetic field is
required to overcome the collective effects of frustration and thermal
fluctuations, and for temperatures higher than $T_3$ we have complete
disorder.

Though to our knowledge the obtained phase diagram has not been presented in
literature before, it has a number of confirmations with experimental
results. In particular it predicts lower value for Neel temperature ($T_1$),
i.e. the temperature at which spontaneous magnetization of subllatice
becomes zero. Fig.5 also indicates the existence of the re-entrant phase
between $T_1$ and $T_2$. Note that for sufficiently strong background
disorder, the measure of which is the ground state entropy, the temperature $%
T_2$ may be equal to zero as in antiferromagnetic Ising model on two
dimensional triangular lattice\cite{A1,P2}.

Peculiar result of our calculation is the range of temperature in between $%
T_1$ and $T_2$, where we have continuous phase transition at a fixed
magnetic field. This is an example of Griffiths singularities well known in
glassy systems\cite{A7,A8}.

Recently we obtain similar singularities in three-site interacting Ising
model on Husimi tree, where magnetization exhibits chaotic size dependence%
\cite{P1}. In three-site interacting Ising model we considered the Lyapunov
exponents as an order parameter and obtained the scaling in terms of
Lyapunov exponents\cite{A9,A10,A11}.

In the presence of Griffiths singularities the scaling relation between
spontaneous magnetization $m_0$ and reduced temperature $t=(T-T_1)/T$ is
well known relation

\begin{equation}
\label{R4}m_0\sim (-t)^\beta 
\end{equation}

when we approach to $T_1$ from low temperature. But the magnetization has
the following non analytic dependence on magnetic field in the temperature
range from $T_1$ to $T_2$:

\begin{equation}
\label{R5}m\sim t^\Phi h^{2t}|\ln h|. 
\end{equation}

For dilute spin glass $\Phi =1+\beta $\cite{A8}.

It will be very interesting to calculate the values of critical exponents $%
\beta $ and $\Phi $ in our case. As to the presence of the ordered state at
finite temperature, one can recall another well known model with
macroscopically degenerate ground state - antiferromagnetic Potts model. It
has been shown in number of work that in some dimensions antiferromagnetic
Potts model can have distinct low-temperature ordered phase with
algebraically decaying correlations \cite{P3}. On the Bethe and Bethe-like
lattice the dimensionality effects can be simulated by changing the
coordination numbers. How this will affect on phase structure we are also
going to find out in our farther work.

This work was partly supported by the Grant-211-5291 YPI of the German
Bundesministerium fur Forshung and Technologie and by the Grant INTAS-96-690.

\begin{figure}[b]
 \epsfysize=12cm\centerline{\epsfbox{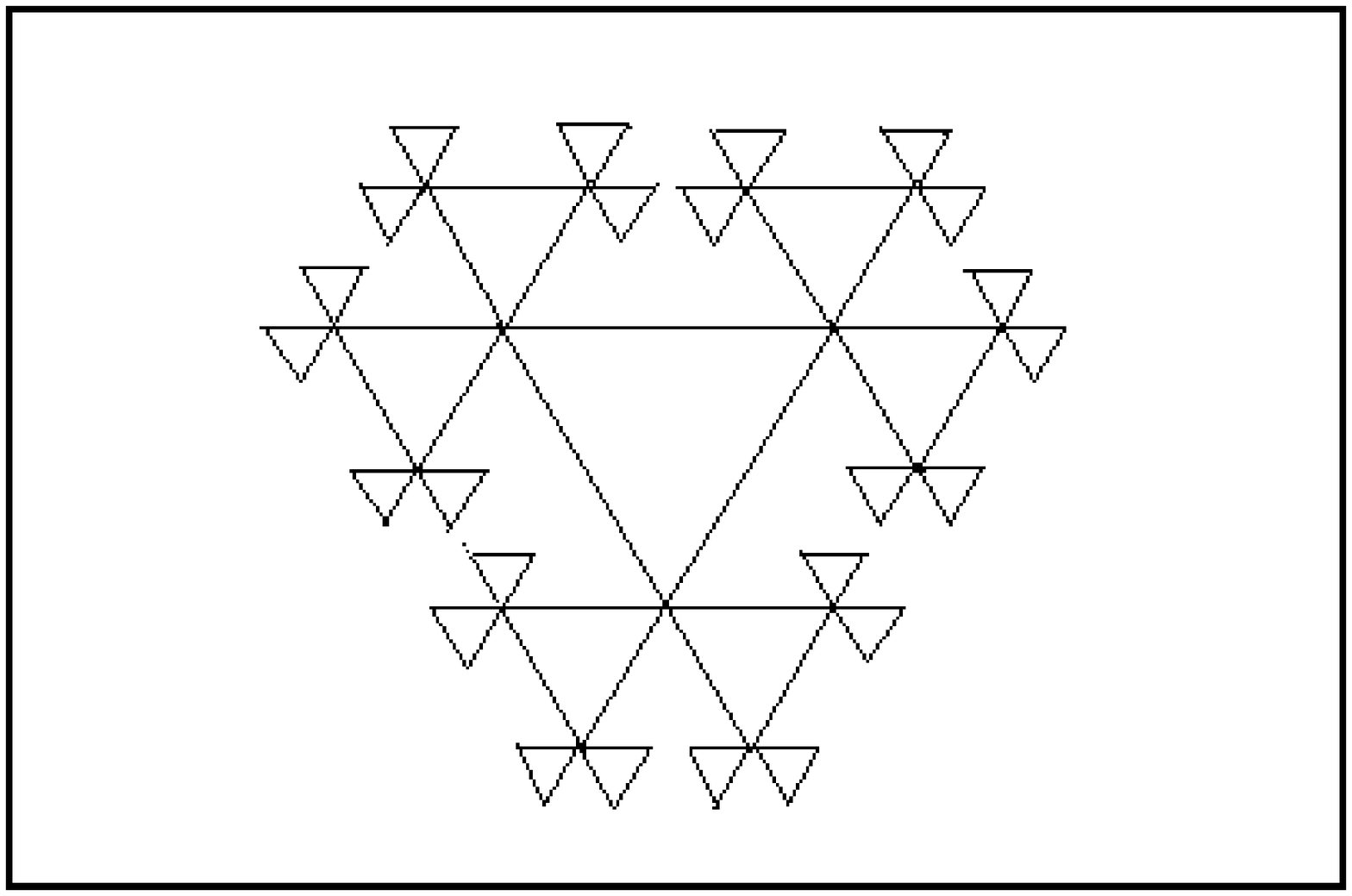}}
 \vspace{6pt}
 \caption{Husimi tree with $\gamma =3$
}
 \label{figone}
 \end{figure}
\newpage
\begin{figure}[b]
 \epsfysize=12cm\centerline{\epsfbox{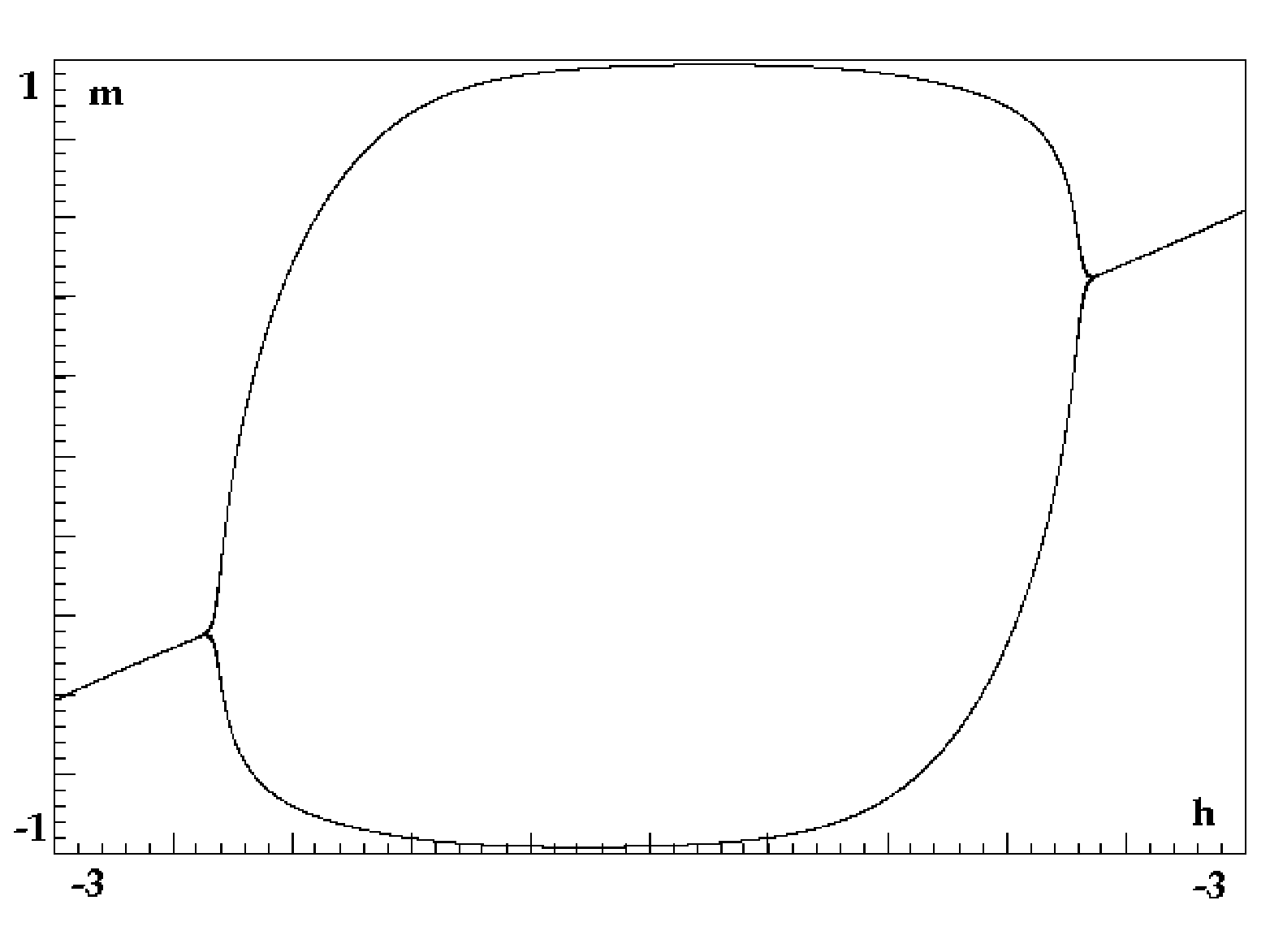}}
 \vspace{6pt}
 \caption{Plot of $m$ versus $h$ ($J=-0.5$,$\gamma =3$)
}
 \label{figtwoa}
 \end{figure}
\newpage
\begin{figure}[b]
 \epsfysize=12cm\centerline{\epsfbox{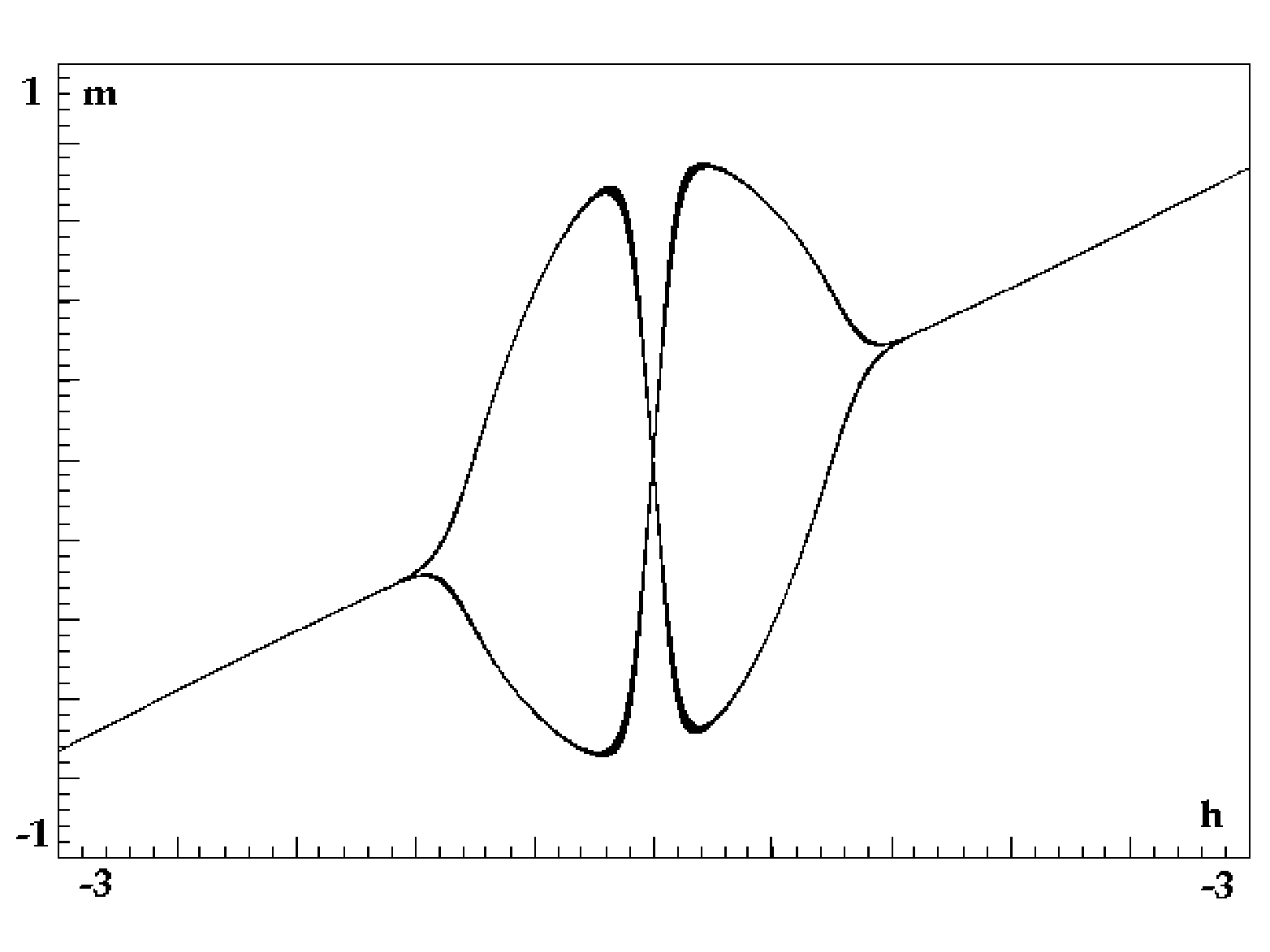}}
 \vspace{6pt}
 \caption{Plot of $m$ versus $h$ ($J=-0.41$,$\gamma =3$)
}
 \label{figtwob}
 \end{figure}
\newpage
\begin{figure}[b]
 \epsfysize=12cm\centerline{\epsfbox{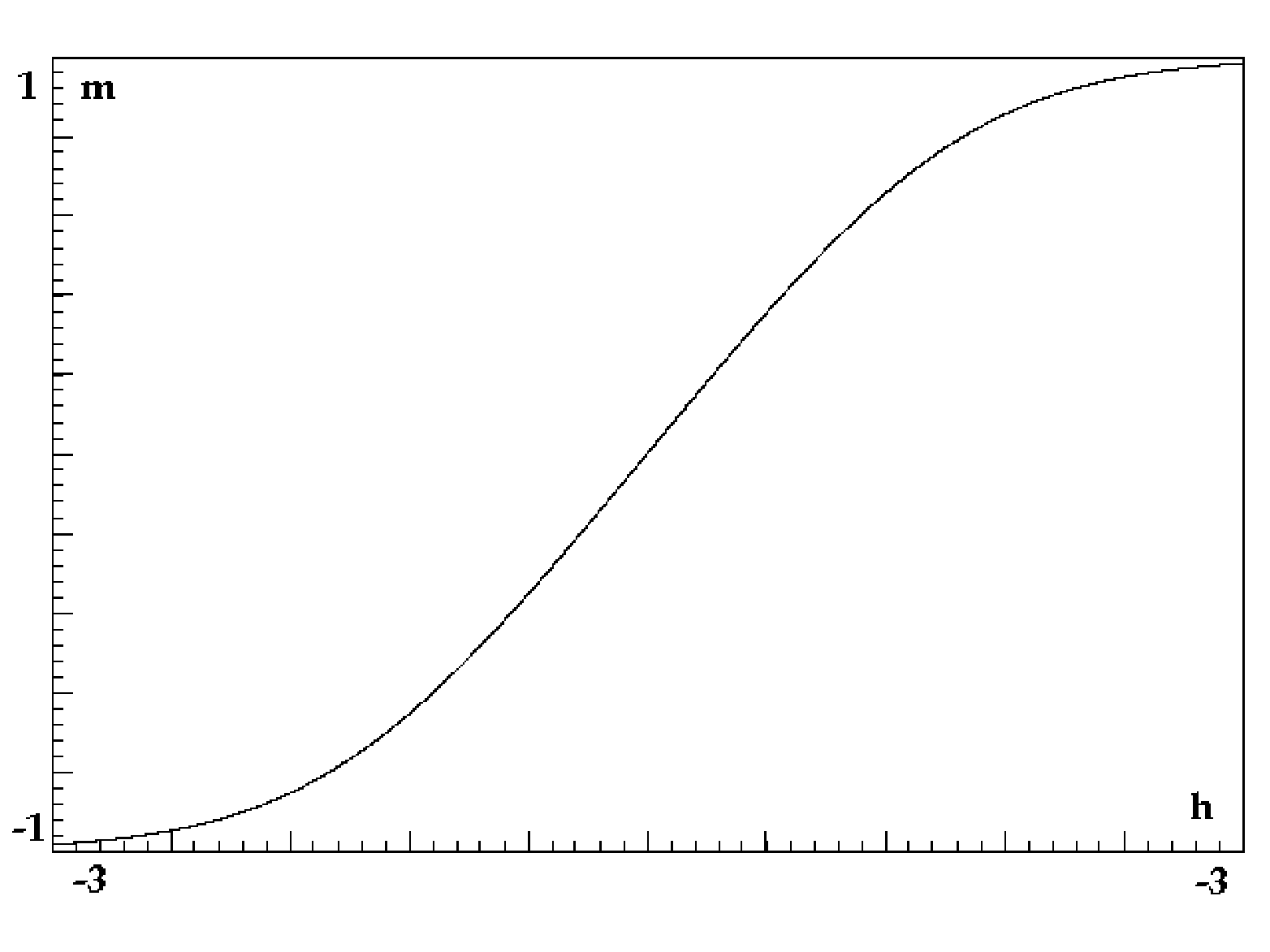}}
 \vspace{6pt}
 \caption{Plot of $m$ versus $h$ ($J=-0.1$,$\gamma =3$)
}
 \label{figtwoc}
 \end{figure}
\newpage
\begin{figure}[b]
 \epsfysize=12cm\centerline{\epsfbox{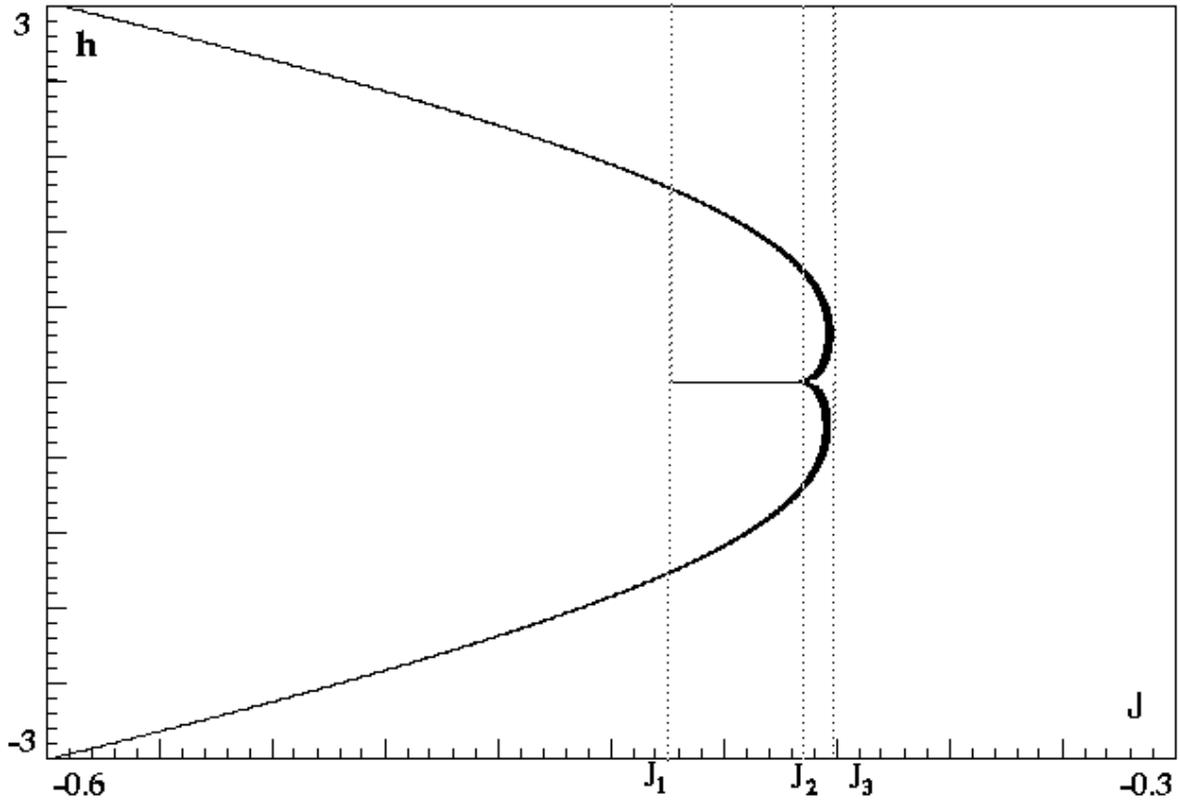}}
 \vspace{6pt}
 \caption{The phase diagramm of antiferromagnetic Ising model in Husimi tree ($\gamma =3$).
}
 \label{figthree}
 \end{figure}

\end{document}